\documentclass[ amsmath, amssymb, aps, 11pt]{revtex4-1}

\usepackage{graphicx}
\usepackage{dcolumn}
\usepackage{bm}
\usepackage{hyperref}
\usepackage{xcolor}

\begin{document}

\title{Dynamically controlled double-well optical potential for colloidal particles}

\author{Thalyta Tavares Martins}
\author{Sérgio Ricardo Muniz}
\affiliation{
Instituto de Física de São Carlos, Universidade de São Paulo, IFSC-USP\\
Caixa Postal 369, CEP 13560-970, São Carlos, SP, Brazil. \\
srmuniz@ifsc.usp.br -- https://orcid.org/0000-0002-8753-4659
}

\begin{abstract}
This preliminary* study presents a simple modulation scheme to dynamically create time-averaged optical potentials to trap colloidal particles using acousto-optical modulation. The method provides access to control experimentally relevant parameters of a tunable double-well potential. We show experimental data successfully adjusting the trapping distance in the range 20--2000 nm and discuss situations arising when reconstructing time-averaged optical potentials from trapped particle data. This is the first step towards studying colloidal particles in dynamically modulated optical potentials to explore the stochastic thermodynamics of mesoscopic systems and small-scale thermo-mechanical machines. \\
(*) \emph{Paper presented at the Conference \href{https://doi.org/10.1109/SBFotonIOPC50774.2021.9461866}{SBFoton-IOPC-2021}.}
\end{abstract}
\maketitle

\section{Introduction}
The ability to optically control particles at the mesoscopic scale, allied to recent advances in light control techniques, has enabled the growth of a new range of applications for optical tweezers in fundamental studies of nonequilibrium  systems. This versatile platform, well-known for studies in biological systems \cite{bennink2001unfolding,shivashankar1997single,bezryadina2016optical}, has been retooled in recent years to study many physical and chemical processes on small scales. Novel theoretical and technological advancements, especially in optical potential engineering, make it possible to explore systems ranging from atoms to nano and microstructures \cite{Marago2013}.

Using acousto-optic modulators (AOM) \cite{martinsaprisionamento,ferrer2015dynamic}, spatial light modulators (SLM) \cite{martin2007design,schonbrun20053d, pedro2021}, digital micromirror devices (DMD) \cite{gauthier2016direct}, and other tools, one can control parameters such as spatial position, intensity, polarization, and the wavefront phase of light.
Combining colloidal particle physics with time-controllable optical potentials has allowed demonstrations of fluctuation theorems \cite{hoang2018experimental,jop2008work,berut2012experimental} and the kinetics of chemical reactions \cite{rondin2017direct}. 

Aiming at a broader range of applications, especially  for fundamental studies, this work proposes a simple but effective scheme to implement dynamical control of optical potentials towards trapping and studying the behavior of colloidal particles in a modulated optical trap using AOM devices. Particularly, we target here the emblematic double-well potential, a paradigmatic model used in many areas. We demonstrate a method to dynamically control the most relevant parameters of this model with ease and precision (down to nanometers). One of our future interests is to use this method to emulate and dynamically control the well-known Kramers model, extensively used in Chemistry and Biophysics.

\section{Methods}

\subsection{Experimental system}
\begin{figure}[tb]
\centerline{\includegraphics[width=11cm]{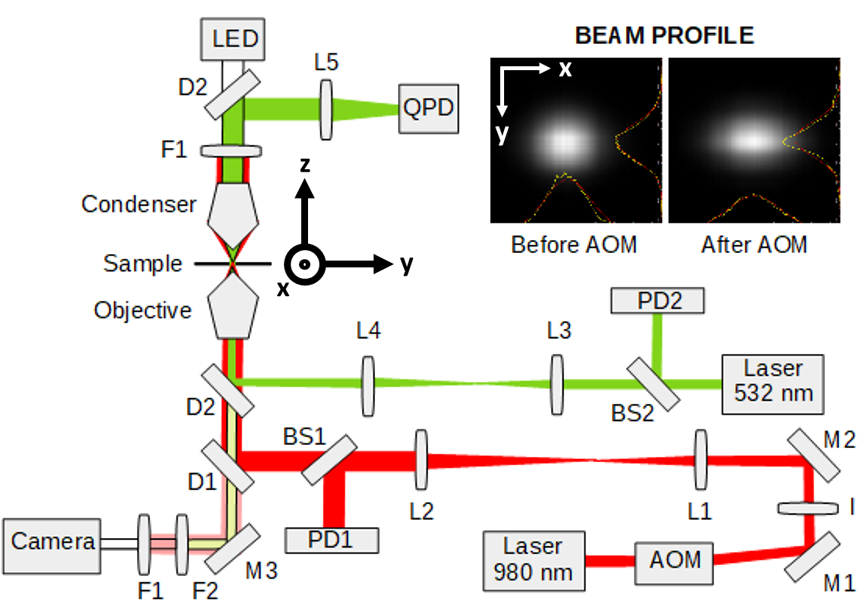}}
\caption{Experimental setup consisting of an infrared laser (980 nm), a green laser (532 nm), an acousto-optical modulator (AOM), objective and condenser, a quadrant photodetector (QPD), a CCD Camera, an illumination withe-LED, photodetectors (PD1-2), mirrors (M1-3), lenses (L1-5), beam splitters (BS1-2), infrared (D1) and green (D2) dichroic mirrors, infrared (F1) and green (F2) filters. Inset shows images of the beam profile of the infrared laser along the $(x,y)$ coordinates, as measured before and after the AOM.}
\label{fig:setup}
\end{figure}

The setup used in this work is shown schematically in Fig. \ref{fig:setup}. It consists of a $980 \ \mathrm{nm}$ fiber-coupled diode laser, followed by an AOM and a telescope before being directed to an optical  microscope.
The home-built microscope has a 100x immersion objective with numerical aperture NA=1.3. It is used to optically trap colloidal particles in solutions kept in a sample holder placed on a 3D nanopositioning stage. 
The sample holder consists of a microscope glass slide with a $10 \ \mathrm{mm}$ diameter and $20 \ \mathrm{\mu m}$ deep reaction well covered by a cover glass. The experiments here used two types of low concentration samples: $2.06 \ \mathrm{\mu m}$ silica beads (Thorlabs OTKBTK) and $1.00 \ \mathrm{\mu m}$ polystyrene beads. The samples were studied in deionized water at room temperature ($300 \ \mathrm{K}$).

Above the sample, a 10x condenser with N.A=0.25 collects the traversing beam and part of the scattered light. Before reaching the microscope objective, the (diffracted first order) beam is magnified by a telescope to overfill the objective entrance slightly. The available power at the sample is approximately $2\,\operatorname{mW}$.

Particle positions are detected using a CCD camera (bright field imaging) and a quadrant photodiode (QPD). The QPD monitors a dedicated low-power probe beam at 532 nm. Red and green dichroic mirrors and filters are used to separate and direct the different wavelengths and avoid saturating the camera. A high-speed multi-function data acquisition (DAQ) board (National Instruments, PCI-6259) controls the AOM and does the data acquisition, reading the QPD and the photodetectors (PDs).

\subsection{Characterization of the trapping potential}
It is possible and frequently desirable to use direct measurements of the trapped particle to determine the effective optical trapping potential. Here, we describe the basic procedure since it is an integral part of the analysis. Later, we show a situation where this procedure does not apply, as it is relevant for this study, pointing out that this well-known method must be used with caution when dealing with dynamically modulated potentials.

In thermal equilibrium with the fluid molecules, the probability distribution of trapped particles follows a Maxwell-Boltzmann distribution, according to
\begin{equation}
    \rho (x) = \rho_0 \exp\Big[-\frac{U(x)}{k_B T}\Big].
    \label{Eq1}
\end{equation}

In (\ref{Eq1}), $U(x)$ represents the trapping potential, $k_B$ is the Boltzmann constant, $T$ is the temperature, and $ \rho_0 $ is a normalization factor such that $ \int \rho (x) dx = 1 $. 

The potential $U(x)$ can be determined from measurements of $\rho(x)$, or measurements of quantities directly proportional to it, following
\begin{equation}
    U(x) = - k_B T \log[\rho(x)]+U_0,
    \label{eq:potencial1}
\end{equation}

\noindent where $ U_0 $ is an arbitrary constant.

One measurement that is relatively easy to obtain experimentally is the position of the colloidal particle as a function of time. Collecting the three QPD voltage outputs, proportional to the particle displacement in $x$, $y$, and $z$ directions, one builds a histogram of positions for each spatial orientation. The calibration constant, used to convert voltage units to displacement units, is obtained moving a trapped particle by a known distance with a 3-axis nanopositioning stage while recording the QPD voltage signal. 

\begin{figure}[tb]
\centerline{\includegraphics[width=11 cm]{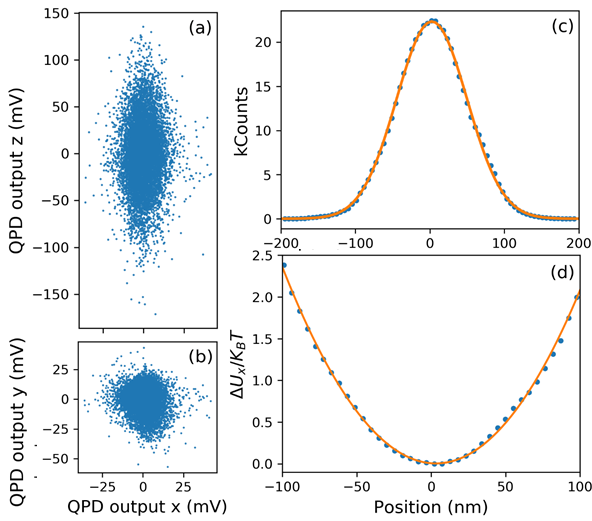}}
\caption{(a)--(b) Point clouds constructed from the QPD readings of positions $(x,z)$ (a), and $(x,y)$ (b), for a trapped particle ($9980$ points collected in $1 \ \mathrm{s}$). (c) Experimental positional histogram in the $x\mathrm{-axis}$. (d) Calculated optical potential, obtained from (c). Data in (c)--(d) were collected at $100\ \mathrm{kHz}$, for $5\ \mathrm{ s}$, with the $2 \ \mathrm{\mu m}$ (silica) particles.}
\label{fig2}
\end{figure}

The position histogram is proportional to $\rho(x)$, at least in equilibrium, and one has the necessary information to effectively measure $U(x)$, directly from this data.
Therefore, by building a normalized histogram of the trapped particle’s positions, one can determine the trapping potential. Figs. \ref{fig2}c and \ref{fig2}d show an example, using experimental data, to illustrate this procedure for the $x$ coordinate.

Since we control (move) the optical potential only along the $x$ direction, using the AOM, all the analysis is presented in terms of this coordinate.
Even though the particle is trapped in three dimensions, as shown in the point clouds constructed from the QPD data in Figs.  \ref{fig2}a and \ref{fig2}b. The AOM causes a slight asymmetry in the Gaussian intensity profile, shown in Fig. \ref{fig:setup}. This effect, however, only makes the trapping a little stronger on the $x\mathrm{-axis}$ than on the $y\mathrm{-axis}$.

The data was collected at a $100 \ \mathrm{kHz}$ rate, still one order of magnitude lower than the bandwidth of the DAQ cards. 

\subsection{Dynamically controlled optical potentials} \label{procedure1}
A focused single laser beam will produce nearly harmonic potential around its equilibrium position, as shown in Fig. \ref{fig2}d. Using an AOM, one can easily control the beam position and the beam's intensity at each position, using an external modulation applied to the frequency and amplitude channels of the AOM driver, respectively.

\begin{figure}[b]
\centerline{\includegraphics[width=12cm]{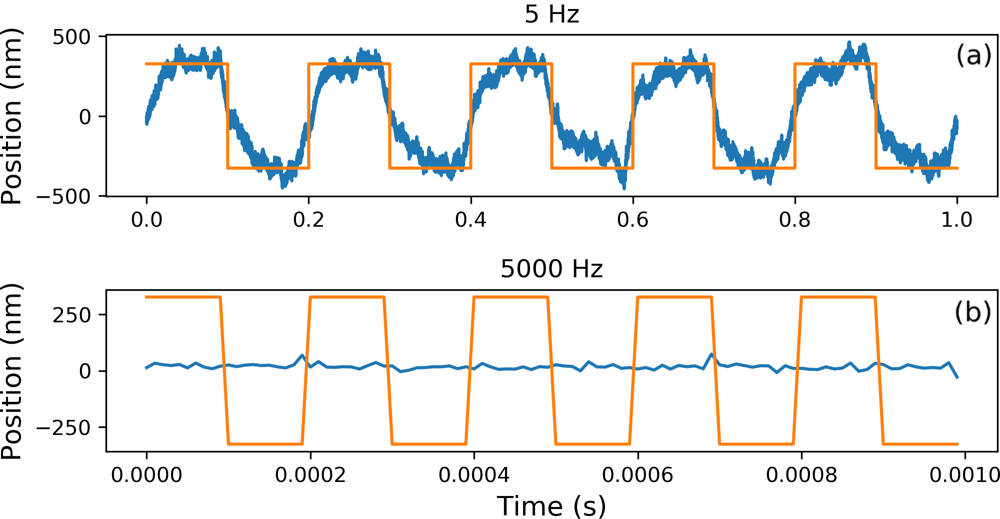}}
\caption{Particle ($2\,\mathrm{\mu m}$) trajectories (blue line) in a dynamically modulated trap, using a square wave applied to the AOM position-channel (in orange), for 5 Hz (a) and 5 kHz (b). The amplitude of peak-to-peak displacement ($A_{PP}$) is $650 \ \mathrm{ nm}$.}
\label{fig:dynamic}
\end{figure}

In this study, we focused our attention on two limits: low switching frequency (5Hz) and high switching frequency (5kHz). Figure  \ref{fig:dynamic} shows data collected at these two limits. The time series for a $2\,\mathrm{\mu m}$ silica bead, subjected to square waveform signal (voltage) applied to the frequency and amplitude channels of the AOMs. The frequency channel allows controlling the trap's position, while the amplitude controls the effective spring constant.

The signal applied in the amplitude channel is responsible for stabilizing the laser power during frequency modulation. The optimal frequency of the AOM is 80 MHz, and at different frequencies, the intensity of the first diffracted order beam drops. Therefore, an amplitude correction is always applied simultaneously with the displacement of the trap to compensate for it.

Fig. \ref{fig:dynamic} shows the modulation applied to the frequency channel (orange lines) and the measured position of the particle (blue lines), synchronously collected with the DAQ board, also controlling the AOM.

The data clearly shows a correlation between the modulation (position of the laser beam) and the position of the trapped particle. At low modulation frequencies, the particle follows the displacement of the optical potential, as expected, with a delay due to the fluid's viscosity. However, at high frequencies, the particle experience the time-averaged potential caused by the fast (5 kHz) switching of the beam's position. The main idea here is to use this feature to propose a simple scheme capable of producing dynamically controlled double-well optical potentials from a time-averaging procedure of a square waveform modulation.

\subsection{Characterization of the static trap}
To determine the control range and the resolution achievable by the proposed method, we performed several position measurements of trapped  particles within a range of displacements of the equilibrium position of the harmonic trap. 

\begin{figure}[tb]
\centerline{\includegraphics[width=10cm]{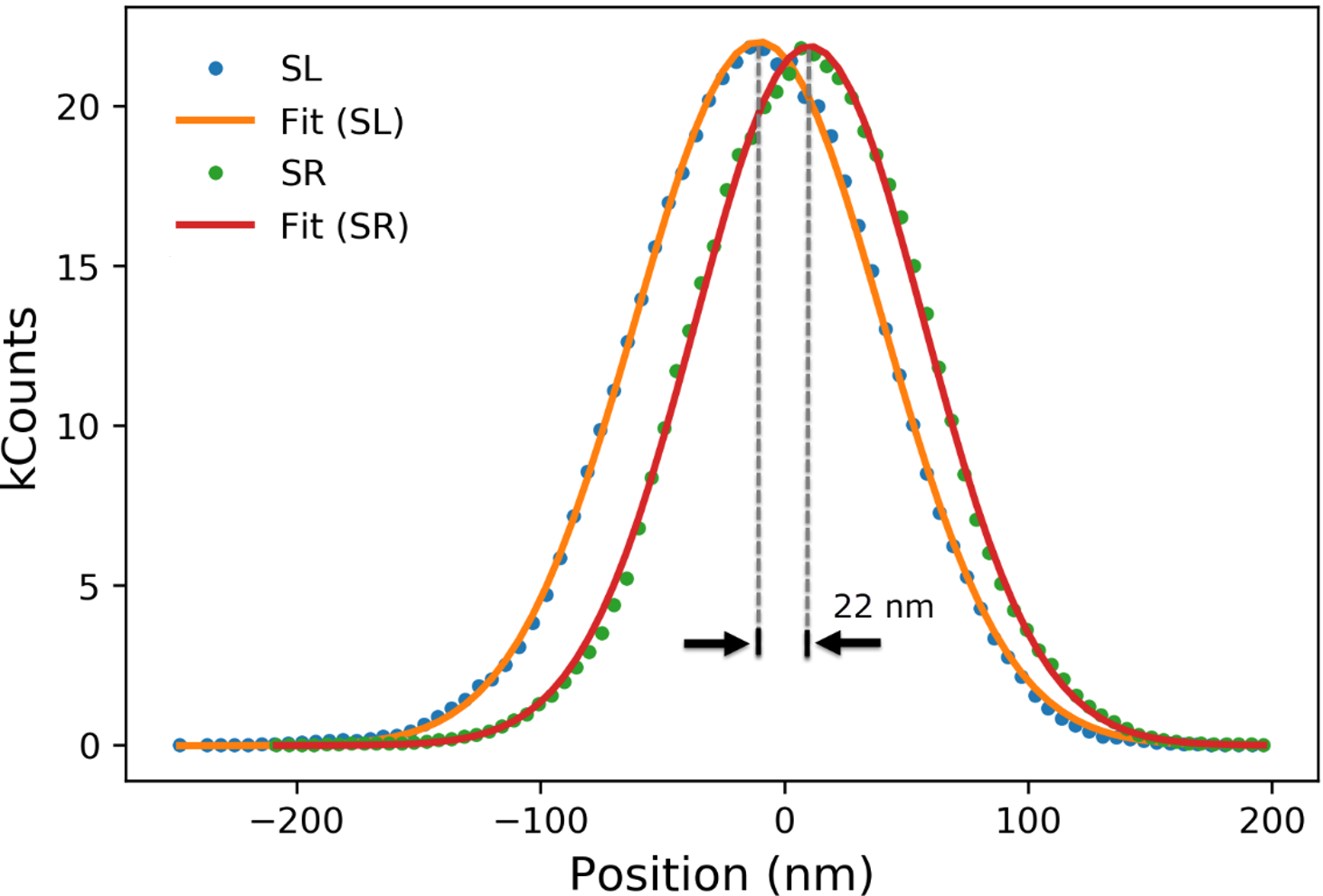}}
\caption{Position histograms of a trapped particle for two static traps (SL and SR) programmed for a displacement of $20 \ \mathrm{ nm}$. The measured value ($22 \ \mathrm{ nm}$) is calculated from the peaks using a Gaussian fit. Data collected at $100\ \mathrm{kHz}$, for $5\ \mathrm{ s}$, with the $2 \ \mathrm{\mu m}$ (silica) particles.}
 \label{fig:static}
\end{figure}

The stochastic trajectories of optically trapped particles were measured, initially in a static optical (single-beam) trap, as shown in Fig. \ref{fig:static}.
Using the procedure described in section \ref{procedure1}, we determined precisely the single-well trap stiffness and the equilibrium position.
The relative error for the spatial separation of the potential wells is around $10 \% $ for distances in the order of hundreds of nanometers, and the percentage is slightly larger when the distance is reduced to a few tens of nanometers.

In Fig. \ref{fig:static}, the histograms of the trajectories of left (SL) and right (SR) static traps indicate an accurate placement of the trap position with $22 \ \mathrm{nm}$ between the traps. A respectable level of control, considering the simplicity of the method. In addition, we can easily observe displacements of up to $2 \ \mathrm{ \mu m}$. The trap displacement is limited by the objective diameter restricting the beam displacement. This highlights the flexibility of the method.

\section{Results}
\begin{figure}[tb]
\centerline{\includegraphics[width=7.6cm]{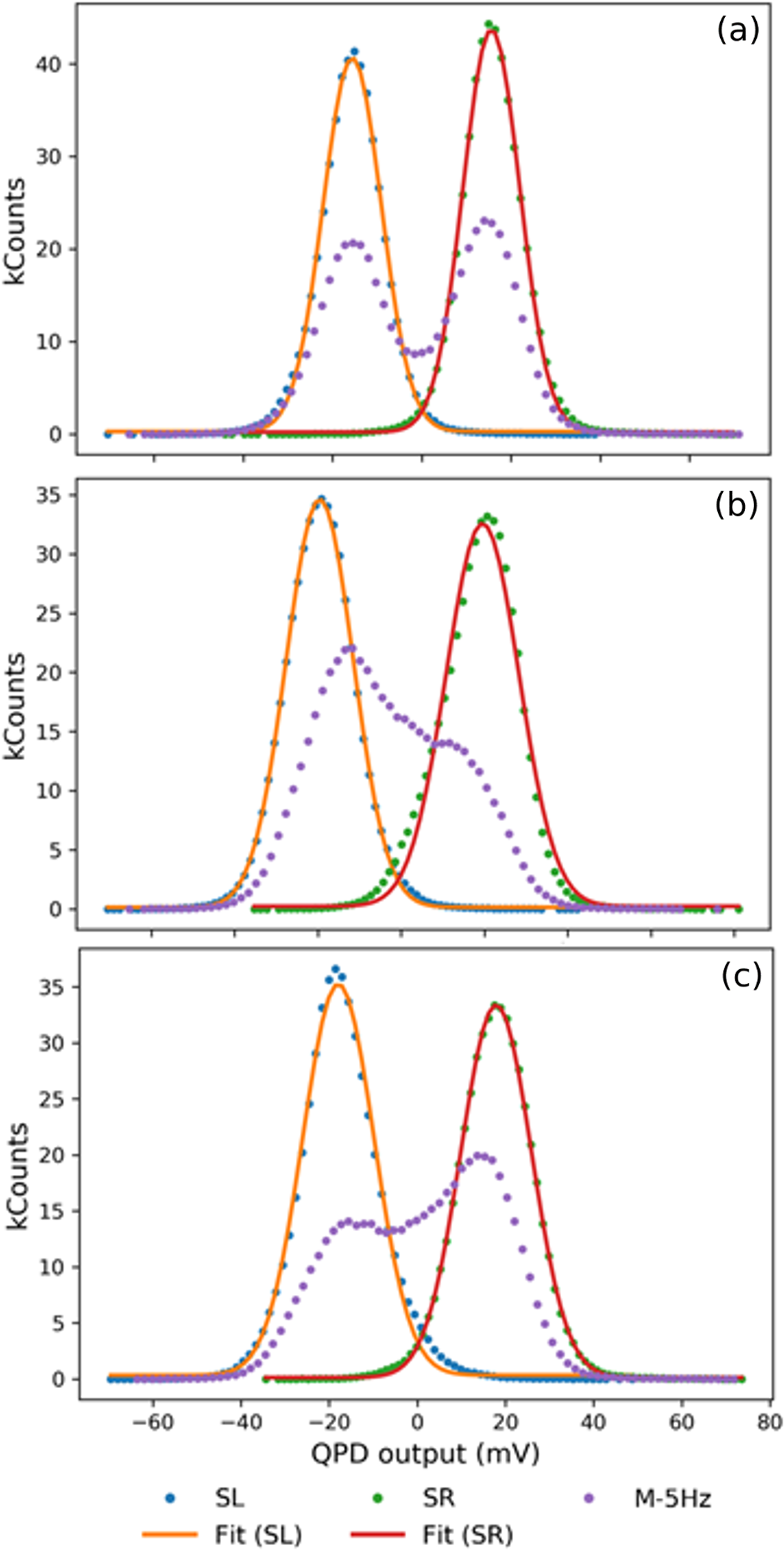}}
\caption{The histograms of two static traps, SL (blue dots) and SR (green dots), with corresponding Gaussian fits, and a modulated trap at 5 Hz (purple dots: M-5Hz) with a displacement of $326 \ \mathrm{nm}$. 
(a) Symmetrical wells ($95\%$ of the laser power in SL and $93\%$ in SR). (b) Asymmetrical wells with the strongest trap in SL ($96\%$ in SL and $48\%$ in SR). (c) Asymmetrical wells with the strongest trap in SR ($48\%$ of the power in SL and $96\%$ in SR). Data collection at $100\ \mathrm{kHz}$, for $5\ \mathrm{ s}$, and with the $2 \ \mathrm{\mu m}$ particles.}
\label{fig:dynamic_5Hz_scal}
\end{figure}

In Fig. \ref{fig:dynamic_5Hz_scal}, the position histograms for SL and SR are shown with the histogram of positions when the beam is modulated at $5 \ \mathrm{Hz}$  ($\mathrm{M-5Hz}$) between the same two static positions, shown as violet-dots.
An additional square wave modulation was applied in the amplitude channel to control the light intensity at each beam position, fine-tuning the trapping constant.

When the laser intensity is the same for both traps (Fig. \ref{fig:dynamic_5Hz_scal}a), the particle spends approximately the same amount of time in each trap. However, for the asymmetric case, the particle spends more time in the well with the greater laser intensity, as shown in Figs. \ref{fig:dynamic_5Hz_scal}b and \ref{fig:dynamic_5Hz_scal}c. This is a direct consequence of the linear relation between the laser intensity and the trap stiffness of the optical potential.

\subsection{Time-averaged optical traps}
This project's long-term goal is to produce double-well optical potentials, dynamically controlling all the model's relevant parameters according to the desired protocol. The idea is to use the time-averaging (i.e., fast-switching) method to create such dynamically controlled double-well potential. Our results, so far, show data consistent with the time-averaged picture, when the light is switched much faster than the particles can follow, typically above $500\,\mathrm{Hz}$. To be safe, we used a modulation of $5\,\operatorname{kHz}$ ($\mathrm{M-5kHz}$).

\begin{figure}[tb]
\centerline{\includegraphics[width=8.5cm]{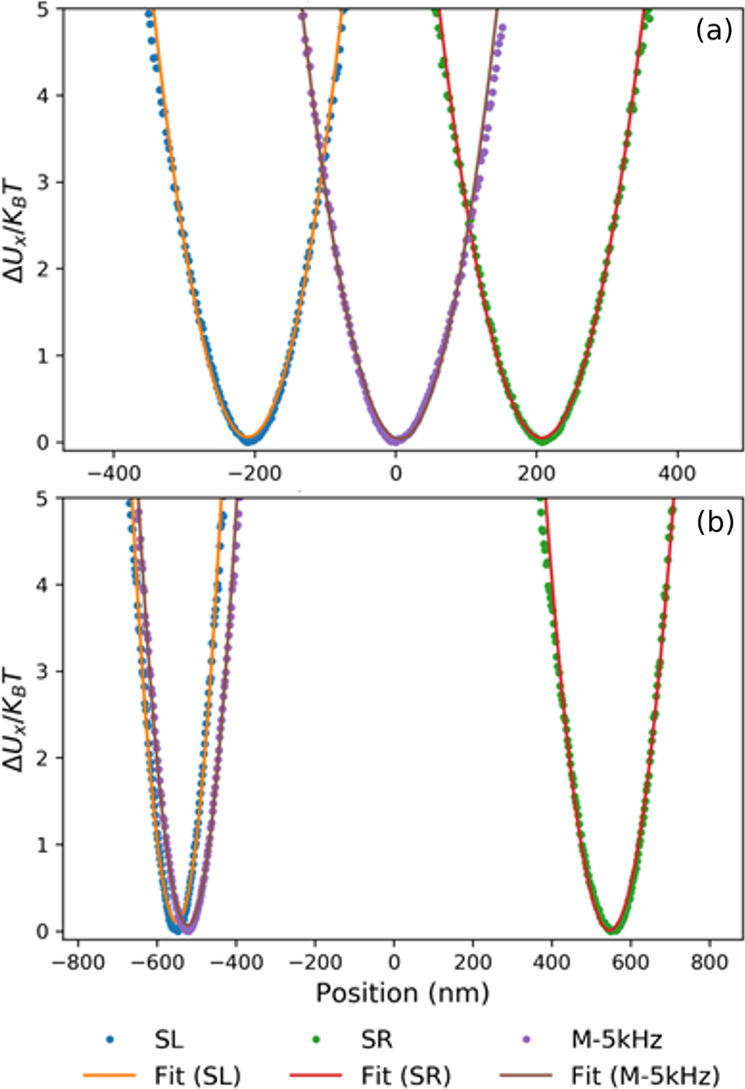}}
\caption{The (quadratic) potential curve of static (SL and SR) and dynamically modulated ($\mathrm{M-5kHz}$) traps. The distances between wells, $A_{PP}$, are $400 \ \mathrm{nm}$ (a) and $1200 \ \mathrm{nm}$ (b). Data collected at $100\ \mathrm{kHz}$ acquisition rate and $5\ \mathrm{ s}$ of measurement time for $1 \ \mathrm{\mu m}$ particles.}
\label{fig:dw}
\end{figure}

For small separations between traps, as in Fig. \ref{fig:dw}a, the optically trapped particle spends most of its time in the region between the two potential minima (measured with the static traps). 
In this intermediate region (between the two static positions), the generated potential is still harmonic, as shown in the reconstructed potential from the position histograms. This behavior is expected, depending on the trap curvature (stiffness) and the beam size (waist).

For large separations, as in Fig. \ref{fig:dw}b, the particle tends to stay trapped in one of the wells, most likely due to a slightly deeper trap in one side or a stochastic fluctuation throwing it in one of the two wells, even when it is perfectly symmetric. If the barrier is sufficiently high, which will be typically the case for well-separated traps, the thermally activated jump is negligible.

It is worth noticing the difference between the reconstructed potential curves at high-frequency (in Fig. \ref{fig:dw}), which is very different from the potential curves one gets at $5\,\mathrm{Hz}$ (in Fig. \ref{fig:dynamic_5Hz_scal}). This happens because of the condition of thermal equilibrium, discussed in section II-B. Therefore, as expected, the reconstructed potential essentially reflects the time spent by the particle at each position, and it does not necessarily represent the actual time-averaged optical potential in nonequilibrium conditions.

\section{Conclusions}
The home-built experimental system allows controlling the trap position from tens of nanometers to a couple of micrometers. This position control, combined with the modulation of the beam intensity, enables the generation of different types of optical potentials by time-averaging the optical potential at high frequencies. We show here that a simple scheme, using a square wave signal, can generate harmonic potentials and multiple traps with a high level of accuracy, but caution must be exercised when trying to reconstruct the optical potential from data when the system is out of equilibrium. Experimental modifications are underway in order to expand range of parameters, and to generate dynamically controllable bistable traps.

\section{Acknowledgment}
We acknowledge the financial support provided by FAPESP (Fundação de Amparo à Pesquisa do Estado de São Paulo), the São Paulo Research Foundation, under research Grants  2019/27471-0 and 2013/07276-1, and also the scholarship funding provided by CAPES and CNPq.




\end{document}